\documentclass[a4paper,fleqn,usenatbib]{mnras}

\usepackage{newtxtext,newtxmath}
\usepackage[T1]{fontenc}
\usepackage{ae,aecompl}
\usepackage{graphicx}	
\usepackage{amsmath}	
\usepackage{amssymb}	

\usepackage[dvipsnames]{xcolor}
\definecolor{mycolor}{rgb}{0.858, 0.388, 0.478}

\title[Gravitational Wave Populations]{Unmodelled Clustering Methods for Gravitational Wave Populations of Compact Binary Mergers}

\author[Jade Powell]{
Jade Powell,$^{1,2}$\thanks{E-mail: jade.powell@ligo.org}
Simon Stevenson,$^{1,2}$
Ilya Mandel,$^{3,2,4}$
Peter Ti$\check{\mathrm{n}}$o$^{5}$
\\
$^{1}$ Centre for Astrophysics and Supercomputing, Swinburne University of Technology, Hawthorn, VIC 3122, Australia \\
$^{2}$ OzGrav: The ARC Centre of Excellence for Gravitational Wave Discovery, Australia \\
$^{3}$ School of Physics and Astronomy, Monash University, Clayton, VIC 3800, Australia \\
$^{4}$ Institute of Gravitational Wave Astronomy and School of Physics and Astronomy,\\
University of Birmingham, Edgbaston, Birmingham B15 2TT, United Kingdom\\
$^{5}$ School of Computer Science, University of Birmingham, Edgbaston, Birmingham B15 2TT, United Kingdom
}

\pubyear{2019}

\begin{document}
\label{firstpage}
\pagerange{\pageref{firstpage}--\pageref{lastpage}}
\maketitle

\begin{abstract}
The mass and spin distributions of compact binary gravitational-wave sources are currently uncertain  due to complicated astrophysics involved in their formation. Multiple sub-populations of compact binaries representing different evolutionary scenarios may be present among sources detected by Advanced LIGO and Advanced Virgo. In addition to hierarchical modelling, unmodelled methods can aid in determining the number of sub-populations and their properties. In this paper, we apply Gaussian mixture model clustering to 1000 simulated gravitational-wave compact binary sources from a mixture of five sub-populations. Using both mass and spin as input parameters, we determine how many binary detections are needed to accurately determine the number of sub-populations and their mass and spin distributions. In the most difficult case that we consider, where two sub-populations have identical mass distributions but differ in their spin, which is poorly constrained by gravitational-wave detections, we find that $\sim 400$ detections are needed before we can identify the correct number of sub-populations.

\end{abstract}


\section{Introduction}
\label{sec:intro}

The Advanced LIGO (aLIGO; \citealp{aLIGO}) and Advanced Virgo (AdVirgo; \citealp{AdVirgo}) gravitational-wave detectors observed ten stellar-mass binary black hole mergers \citep{o1cbcsearch, 2018arXiv181112907T} and a binary neutron star merger \citep{PhysRevLett.119.161101} during the first two observing runs. The third observing run is currently underway, and several gravitational-wave detection candidates have been identified and circulated to electromagnetic partners via GCN\footnote{\url{https://gcn.gsfc.nasa.gov}}.
These merging compact binaries may have formed through several different formation mechanisms. Some of the possible formation mechanisms are classical isolated binary evolution \citep[e.g.,][]{2016Natur.534..512B}, dynamical interactions in dense stellar environments \citep[e.g.,][]{10.1093/mnras/173.3.729}, chemically homogeneous evolution \citep{2016MNRAS.458.2634M, 2016A&A...588A..50M}, triple system formation \citep[e.g.,][]{Antonini:2017ash, Rodriguez:2018jqu}, or even a single event gravitationally lensed into multiple events \citep{2019arXiv190103190B}. Multiple compact binary gravitational-wave detections enable the study of the properties of populations of sources, which will constrain formation mechanisms \citep[e.g.,][]{2017Natur.548..426F, 2017MNRAS.471.2801S, 2015ApJ...810...58S, 0264-9381-34-3-03LT01, 2017PhRvD..95l4046G, 2018arXiv180102699T, 2018arXiv180506442W, 2017ApJ...846...82Z}. 

The rates, masses and spins of binary black holes can inform our understanding of their formation. If reliable models are available for different formation mechanisms then hierarchical inference can be applied to compact objects detected in gravitational waves.  This approach is optimal when models can be trusted, and allows the mixing ratios of different sub-populations and the properties of each population (e.g., the physics governing natal kicks received during core collapse or common envelope phases during binary evolution) to be measured \citep{2015PhRvD..91b3005F}.

Several studies have compared mass distributions to population synthesis models \citep{2015ApJ...810...58S, 2017ApJ...846...82Z, 2018PhRvD..97d3014W, 2018PhRvD..98h4036G,TaylorGerosa:2018}.  For example, \citet{2018MNRAS.477.4685B} argued that the parameters describing common envelope ejection efficiency, stellar wind strength and natal kicks can all be measured to an accuracy of a few percent with a thousand detections provided the parametrised evolutionary model is accurate. 

The spin magnitudes and the alignment between the black hole spins and the orbital angular momentum carry information about the black hole binary formation mechanism.  However, spin is difficult to constrain with gravitational-wave observations. Isotropic spin-orbit misalignment angles are expected for dynamically formed binaries \citep{10.1093/mnras/173.3.729, 2007ApJ...661L.147B, 2016ApJ...832L...2R}, but isolated binaries are expected to be preferentially aligned with the orbital angular momentum \citep[e.g.,][]{2016Natur.534..512B,Kushnir:2016,Qin:2019}. To determine the distinguishability of these two formation mechanisms with gravitational-wave observations, hierarchical models have been applied to real and simulated gravitational-wave measurements of spin-orbit misalignment angles \citep{2017MNRAS.471.2801S, 0264-9381-34-3-03LT01, 2017PhRvD..96b3012T, 2017Natur.548..426F, 2018arXiv181112940T, 2018ApJ...854L...9F}.

However, given the many modelling uncertainties, unmodelled or weakly modelled inference is a necessary back-up tool for studying source populations.  Previous studies include fitting phenomenological population hyper-parameters to mass distributions, assuming the mass distribution is a power-law inherited from the stellar initial mass function \citep{PhysRevD.95.103010, 2018arXiv180506442W, 2019MNRAS.484.4216R, 2018arXiv181112940T}, and including an upper mass gap and an excess of black holes near $40\,M_{\odot}$ \citep{2017ApJ...851L..25F, 2018arXiv180102699T}.

Unmodelled clustering techniques are particularly useful for interpreting population data when there is limited confidence in the available models.  \citet{2015MNRAS.450L..85M} argued that clustering can distinguish mock sub-populations of binary neutron stars, neutron star black holes, and binary black holes with tens of observations using only information about the masses of the two compact objects.
 \citet{2017MNRAS.465.3254M} demonstrated that this is achievable for populations of compact binary mergers whose true mass distributions do not overlap (but whose measured properties do overlap because of measurement uncertainty).  The method used in \citet{2017MNRAS.465.3254M} involves reconstructing the observed mass distribution of merging compact binaries with a Gaussian process prior over a pixellated two-dimensional mass distribution with $\sim 10^{2}$ bins.  Clusters are found in the reconstructed mass distribution with a `water filling' algorithm \citep{2017MNRAS.465.3254M}.  
 
The clustering method of \citet{2017MNRAS.465.3254M} scales poorly to a high-dimensional observable parameter space.  The number of bins is exponential in the number of dimensions; $\sim 10^4$ bins will be needed in 4 dimensions.  The `water filling' clustering algorithm does not scale trivially to higher dimensions.

A Gaussian mixture model is an alternative clustering algorithm that scales well with the number of dimensions.  A Gaussian mixture model fits a convex combination of multivariate Gaussian distributions to the input data. \citet{2017arXiv171202643W} apply a Gaussian mixture model to two simple examples. The first is a synthetic population of 30 binaries containing two sub-populations widely separated in the parameter space. One of the sub-populations has high mass and low spin, and the other has low mass and high spin. They predict the correct number of sub-populations but incorrectly predict the features of each sub-population when there are only 30 detections. The second example is 1000 perfect measurements from a single power-law distribution containing a mass gap. 

Gaussian mixture models have also been applied to gravitational-wave detector noise transients, as finding different populations of noise transients can help identify their origin \citep{2015CQGra..32u5012P}.  

We expand on previous studies by applying clustering methods to sources sampled from a mixture of five complex sub-populations of simulated compact binaries.  We model the imperfect inference on  individual source parameters from their gravitational-wave signatures via realistic simulated posterior distributions. We apply a Gaussian mixture model \citep{scikit-learn} assuming that the source population can be represented by a mixture of multivariate Gaussians. We apply this method simultaneously to both the mass and spin parameters for all of the binary systems considered. We consider how well our method can predict the correct properties of each sub-population. We show that the freedom in the choice of parametrisation, the choice of cluster shape, and the choice of a distance metric can have a large effect on the results. 

This paper is structured as follows: Section \ref{sec:pops} describes the mock astrophysical sub-populations considered in this study. Section \ref{sec:posts} explains how the mock inferred posteriors on the gravitational-wave parameters are produced. Section \ref{sec:method} presents the Gaussian mixture model method applied to the simulated sub-populations. The results are shown in Section \ref{sec:results}, and a conclusion and discussion are given in Section \ref{sec:conclusion}.

\section{Sub-populations}
\label{sec:pops}

\begin{figure*}
\centering
\includegraphics[width=0.32\textwidth]{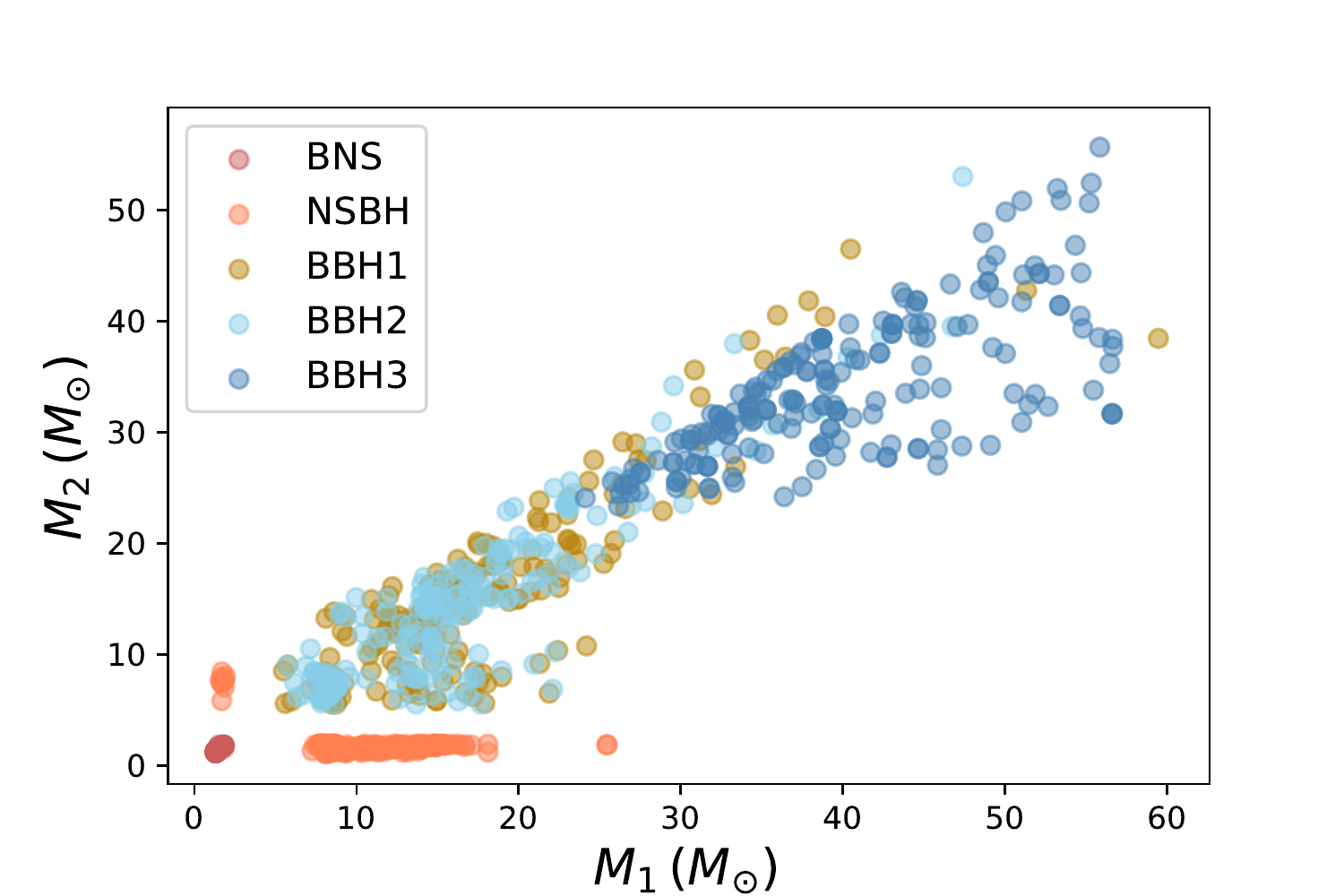}
\includegraphics[width=0.32\textwidth]{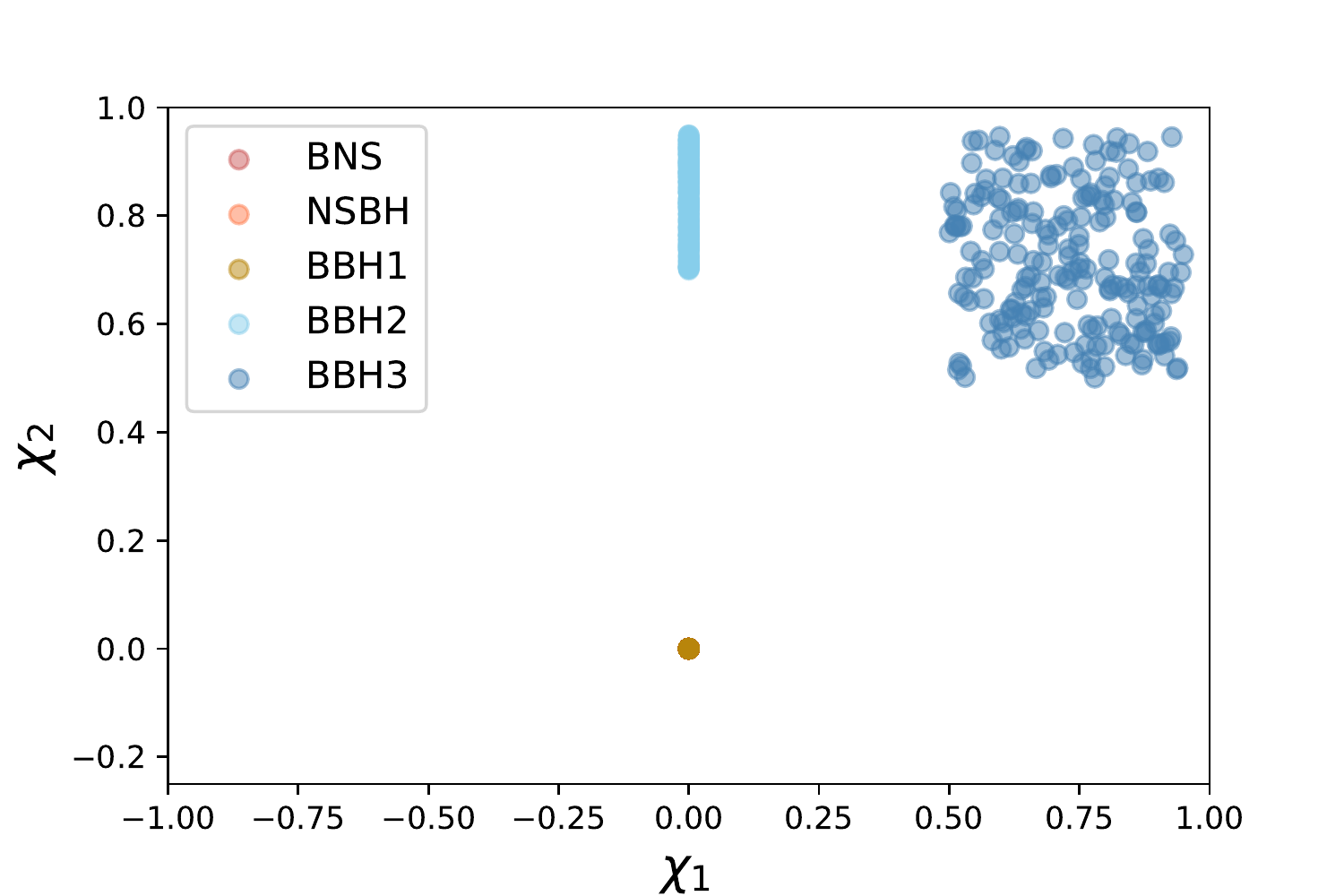}
\includegraphics[width=0.32\textwidth]{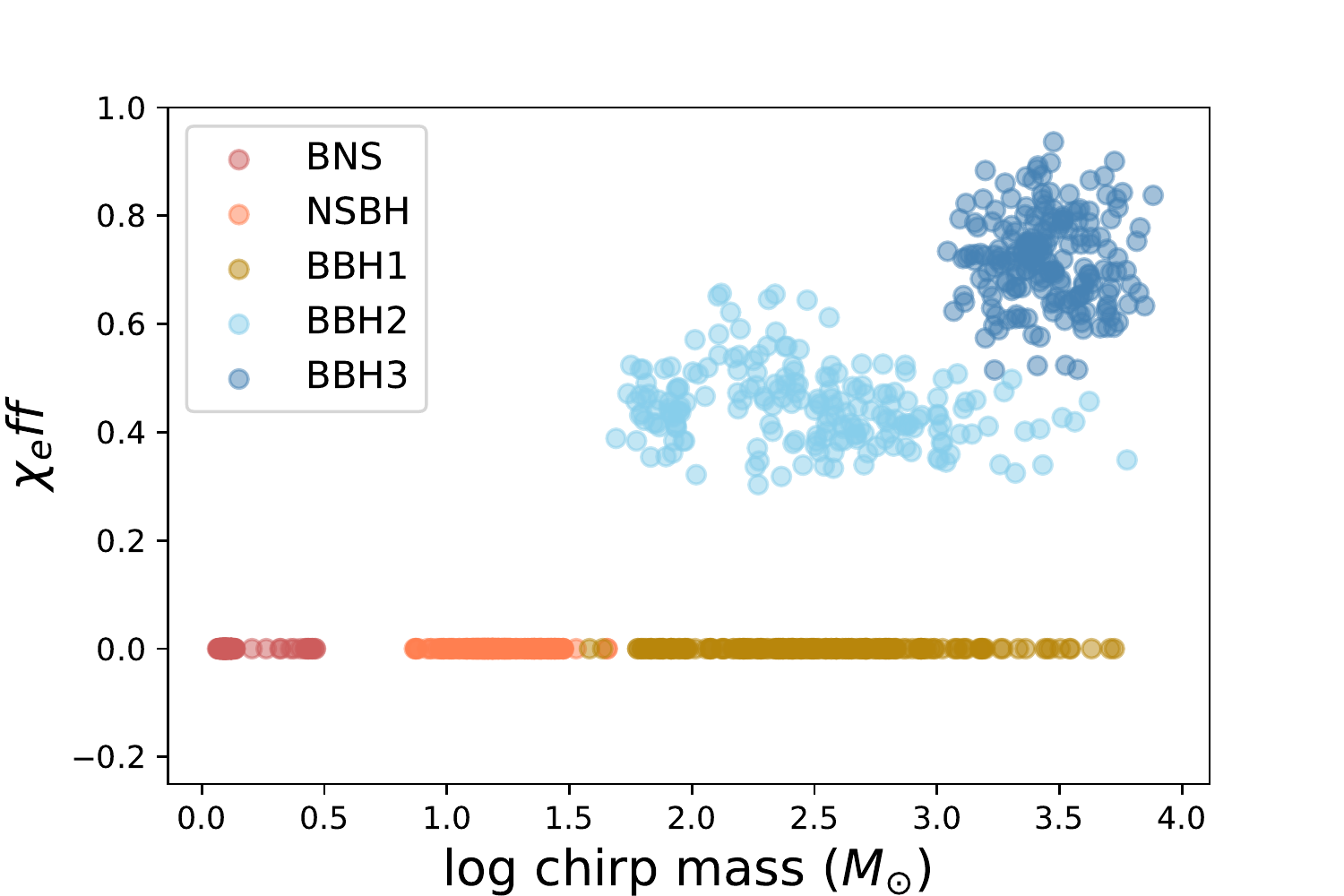}
\caption{The true mass and spin values of 1000 simulated compact binaries. The different colours correspond to the different sub-populations. (left) The true values of the individual masses.  Populations BBH1 and BBH2 have identical mass distributions. (middle) The true values of the aligned spins of the black hole sub-populations. The sub-populations containing neutron stars all have zero spin. (right) The true values converted to effective spin and log chirp mass. }
\label{fig:pops}
\end{figure*}

\begin{table*}
\begin{center}
\begin{tabular}{|c|c|c|c|c|c|}
\hline
short name & source type & mass distribution & $m_1$ 90\% range & $m_2$ 90\% range & spin distribution \\  \hline 
 BNS  & neutron star binaries & \citet{2015ApJ...806..263D}    & $1.3\,M_{\odot}$ - $1.32\,M_{\odot}$ & $1.2\,M_{\odot}$ - $1.3\,M_{\odot}$ & $\chi_{1} = \chi_{2}=0$  \\
 NSBH & NS--BH binaries & \citet{2015ApJ...806..263D} & $7\,M_{\odot}$ - $15\,M_{\odot}$  & $1.3\,M_{\odot}$ - $1.9\,M_{\odot}$ & $\chi_{1} = \chi_{2}=0$  \\ 
 BBH1 & black hole binaries    & \citet{2015ApJ...806..263D}   & $7\,M_{\odot}$ - $28\,M_{\odot}$  & $5\,M_{\odot}$ - $24\,M_{\odot}$ & $\chi_{1} = \chi_{2}=0$  \\
 BBH2 & black hole binaries    & \citet{2015ApJ...806..263D}   & $7\,M_{\odot}$ - $26\,M_{\odot}$  & $5\,M_{\odot}$ - $23\,M_{\odot}$ & $\chi_{1} = 0$, $\chi_{2} = U(0.7, 0.95)$   \\ 
 BBH3 & black hole binaries   & \citet{2016MNRAS.458.2634M}    & $29\,M_{\odot}$ - $56\,M_{\odot}$   & $25\,M_{\odot}$ - $44\,M_{\odot}$ & $\chi_{1} = \chi_{2} = U(0.5, 1.0)$   \\  \hline
\end{tabular}
\caption{Details of the 5 sub-populations considered in this study. The first is non-spinning binary neutron stars. The second is neutron star black hole binaries. The third and fourth sub-populations are black holes with the same mass distribution but different spin distributions. The fifth population contains higher mass black holes with larger spins. }
\label{tab:pops}
\end{center}
\end{table*}

In this study, we simulate five sub-populations of compact binaries. There are a total of 1000 simulated binaries with $20\%$ from each sub-population. The choice of $20\%$ is arbitrary as we do not know what the true mixture fraction of sub-populations will be. This choice is not intended to be realistic: e.g., it does not respect the currently observed ratios between different binary types \citep{2018arXiv181112907T}. In section \ref{sec:results}, we show that varying the mixing ratios does not significantly affect our results.

We consider 4 parameters for each simulated binary. They are the companion masses $m_{1}$ and $m_{2}$, and the aligned spins $\chi_{1}$ and $\chi_{2}$, where 
\begin{equation}
\chi_{1,2} = a_{1,2} \cos\theta_{1,2}.
\end{equation}
The spin-orbit misalignment angle is given by $\theta$, and $a$ is the spin magnitude.

The true masses and spins of the compact objects in each sub-population are shown in Figure \ref{fig:pops}, and further details of their ranges are given in Table \ref{tab:pops}. We consider our sub-populations as toy models that demonstrate our method. 

The first three sub-populations are binary neutron stars (BNS), neutron star black hole binaries (NSBH), and binary black holes (BBH1).  The mass distributions are taken from \citet{2015ApJ...806..263D} as in the previous studies of \citet{2015MNRAS.450L..85M, 2017MNRAS.465.3254M}. The sub-populations were produced by the population synthesis code \texttt{StarTrack} \citep{2008ApJS..174..223B} and down-selected to only include binaries in the aLIGO and AdVirgo detection range.  Figure \ref{fig:pops} shows the clear gap in mass between the three sub-populations.  We further assume that all compact objects in these three populations have zero spin.

The fourth sub-population (BBH2) has the same mass distribution as BBH1 but allows for the secondary companion to have aligned spin: $a_{1}$ is zero, and the secondary black hole spin $a_{2}$ is uniformly distributed between 0.7 and 0.95. 
For binary black holes formed through common envelope evolution, the first born black hole is likely to  have negligible spin \citep{2018A&A...616A..28Q,Bavera:2019}. This is because most of its angular momentum is stored in the envelope during the giant stage, and is removed during mass transfer or common envelope evolution. The spin of the second born black hole is determined by the effects of wind mass loss and tides on its progenitor, a helium star \citep{2016MNRAS.462..844K,2018MNRAS.473.4174Z,2018PhRvD..98h4036G,2018A&A...616A..28Q,Bavera:2019}. If the binary has a sufficiently short orbital period, the helium star may be spun up through tides and form a rapidly spinning black hole. Otherwise, in binaries with larger orbital periods, the helium star is not spun up through tides, and loses angular momentum through stellar winds. We use this to motivate the spin values of BBH1 and BBH2. Distinguishing between BBH1 and BBH2 is expected to be the most difficult example considered in this study due to the poor constraints on gravitational-wave spin measurements.

Binary black holes formed through binary evolution are expected to have their spins preferentially aligned with the orbital angular momentum.  The spins are perfectly aligned in both sub-populations that have spins, i.e.,  $\theta = 0$ for all our binary systems. Misalignments are fairly uncertain, due to uncertainties both in the magnitude of black hole kicks, and in realignment processes through binary evolution \citep{2018PhRvD..98h4036G, 2018PhRvD..97d3014W}. We do not expect the assumption of spin-orbit alignment to have a significant effect on our results as the distributions would still be reasonably well separated if a more moderate amount of misalignment was included, as expected in more realistic models.

The fifth sub-population (BBH3) consists of black holes with a higher mass distribution based on the results of \citet{2016MNRAS.458.2634M}, where they consider merging binary black holes formed through chemically homogeneous evolution in short period stellar binaries. The individual masses are distributed between approximately $25\,M_{\odot}$ and $50\,M_{\odot}$. The aligned spins of the individual black holes are motivated by the results of \citet{2016A&A...588A..50M} and are uniformly distributed between $\chi = 0.5$ and $\chi = 1.0$.

\section{Posteriors}
\label{sec:posts}

\begin{figure*}
\centering
\includegraphics[width=0.3\textwidth]{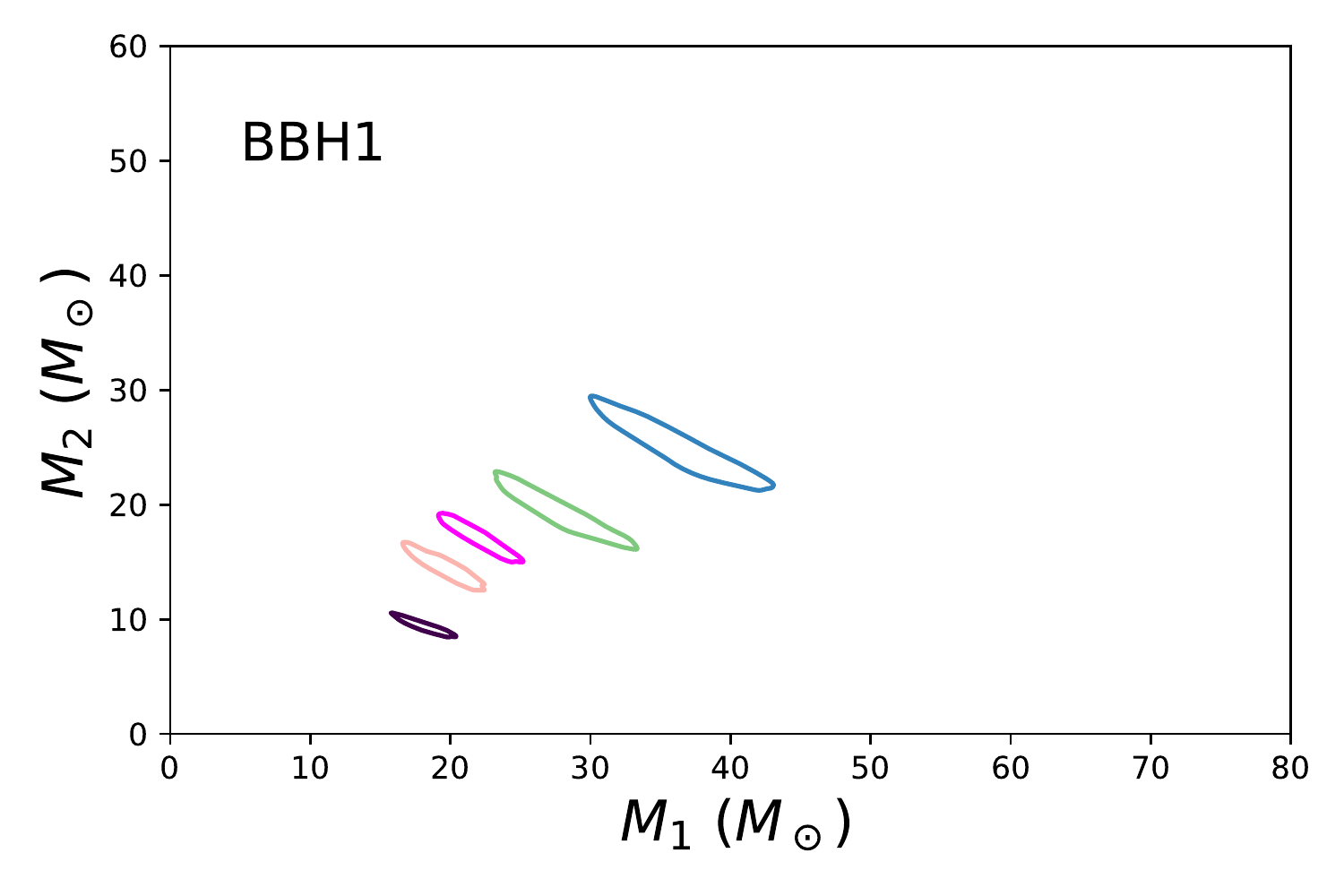} 
\includegraphics[width=0.3\textwidth]{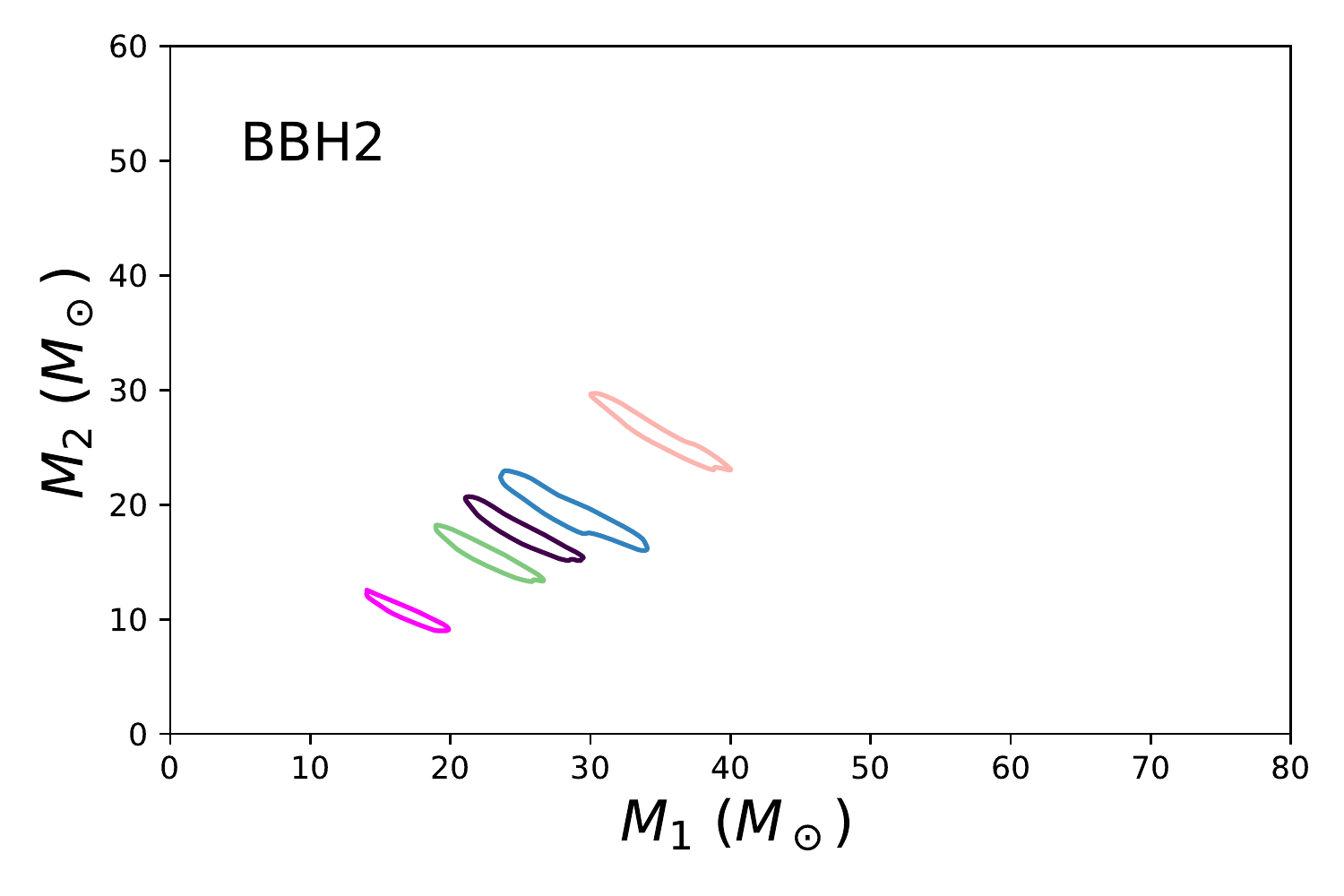} 
\includegraphics[width=0.3\textwidth]{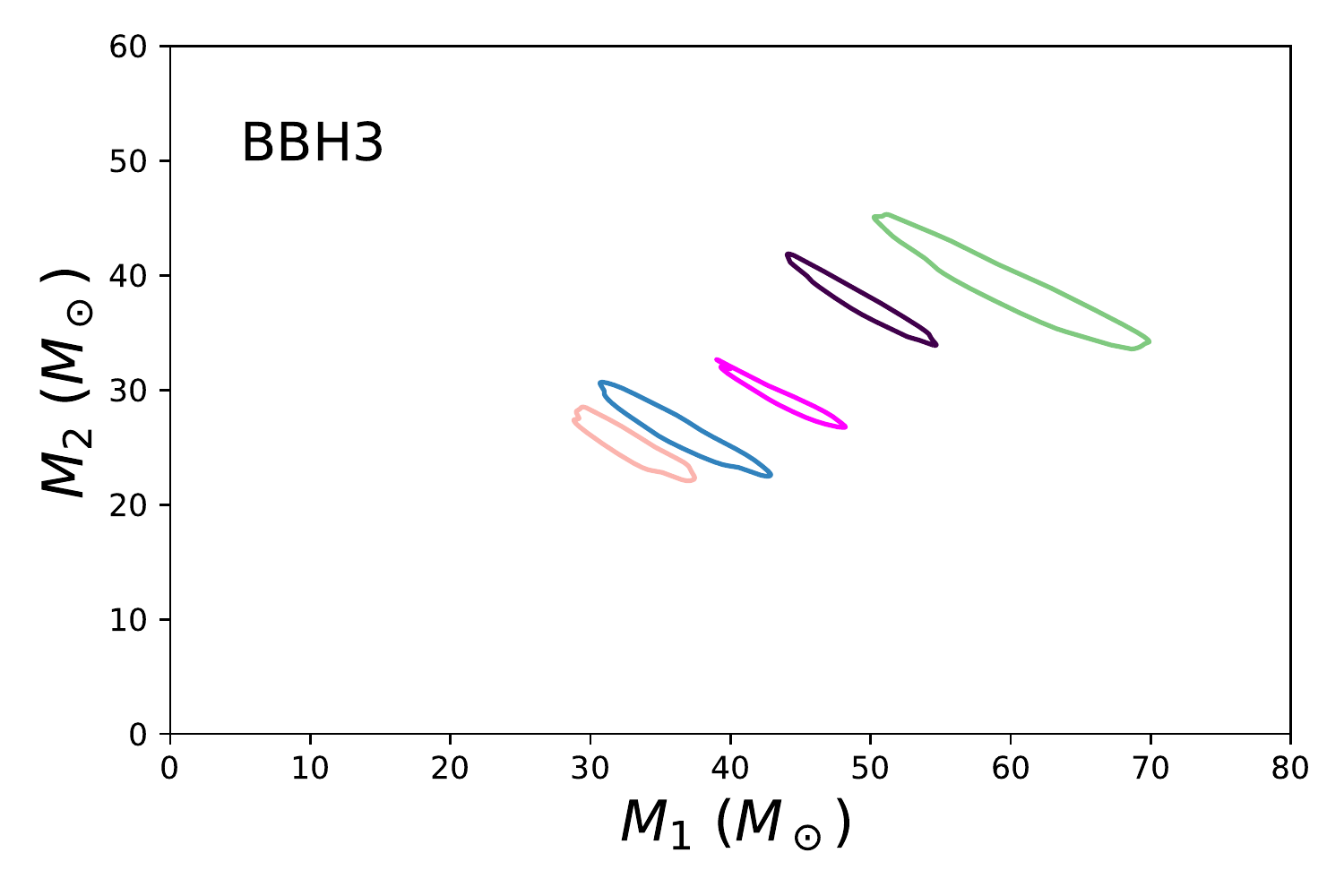} 
\includegraphics[width=0.3\textwidth]{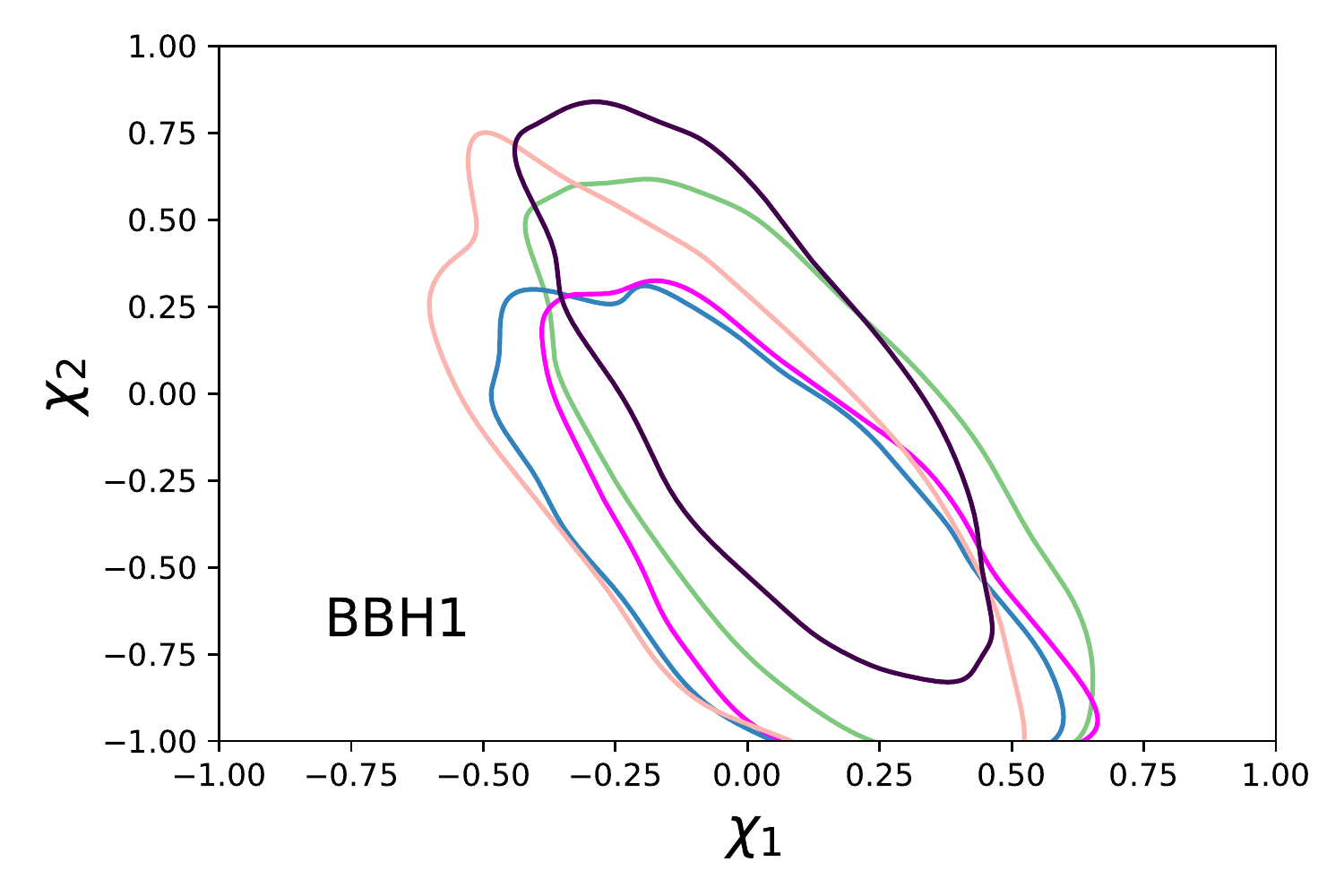} 
\includegraphics[width=0.3\textwidth]{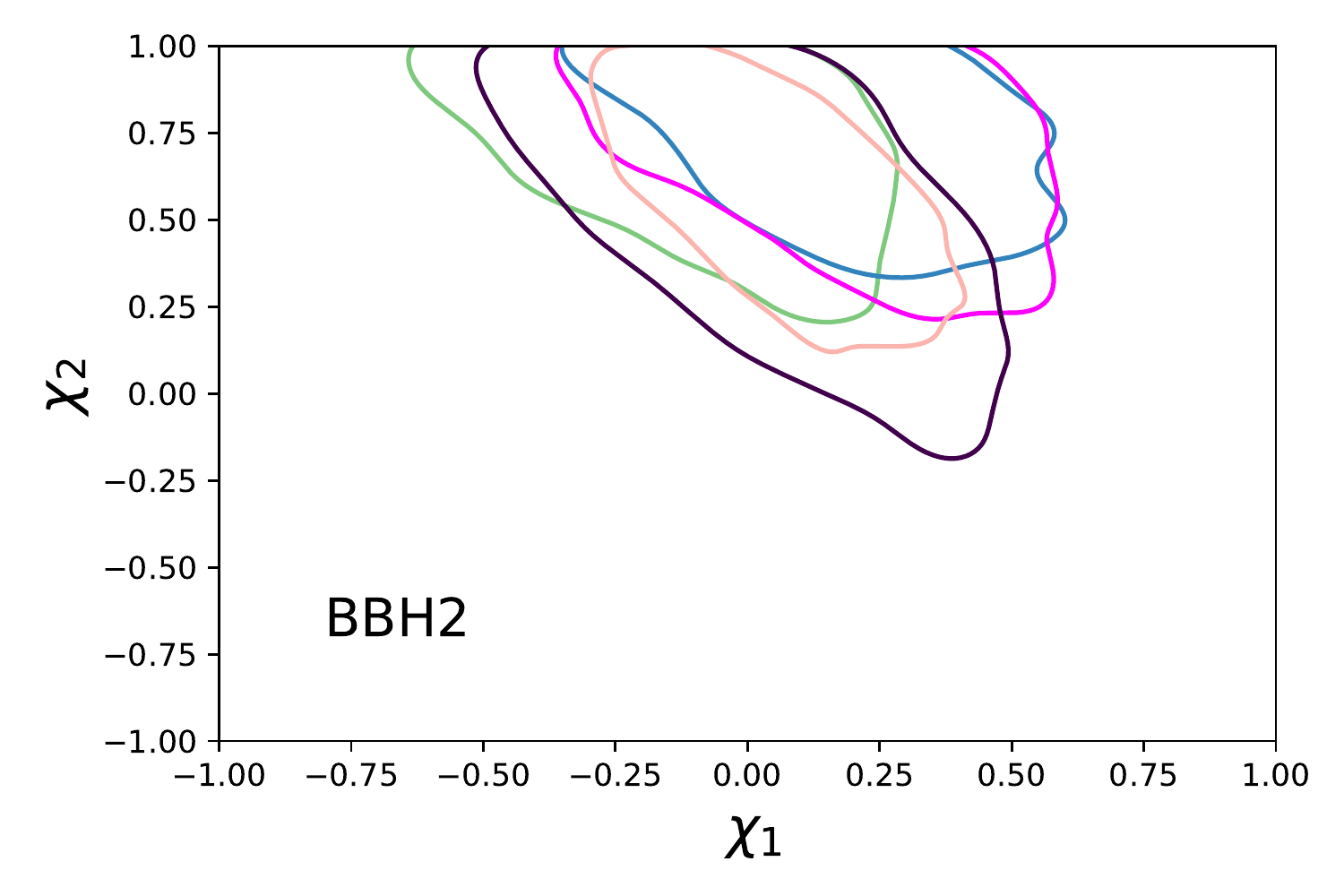} 
\includegraphics[width=0.3\textwidth]{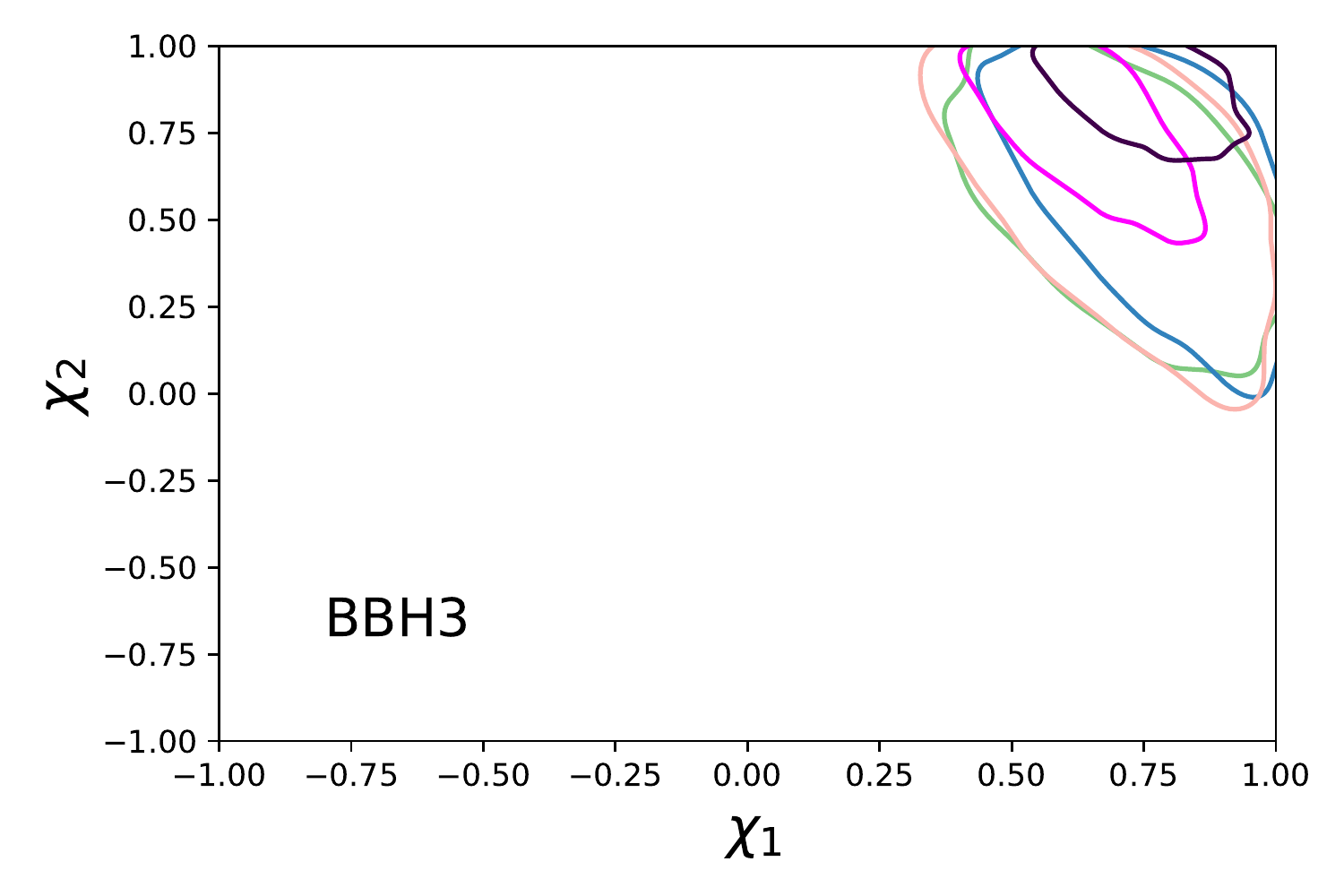}   
\caption{Five representative 90\% confidence interval contours for the mass and spin posteriors of the three different binary black hole populations.  The mass posteriors show the typical banana shape and are better constrained than the spin. BBH1 and BBH2 have the same mass distributions but differ in their aligned spins. }
\label{fig:posts}
\end{figure*}

Bayesian parameter estimation is applied to data containing detected gravitational-wave signals to produce posterior distributions for the astrophysical parameters of the source \citep{2015PhRvD..91d2003V}. For compact binary signals, the parameters include the mass, spin, eccentricity, distance, inclination, and sky position. Bayesian inference is computationally expensive for a large number of detections. Therefore, 
for this study, we produce realistic artificial posteriors for mass parameters as in \citet{2017MNRAS.465.3254M}, and spin parameters as in \citet{2017MNRAS.471.2801S}.

For the mass posteriors, we first sample values for $m_1$ and $m_2$ from each of the 5 sub-populations described in Section \ref{sec:pops}. For each binary system, we then draw a signal to noise ratio (SNR) $\rho$ value from a $p(\rho) \propto \rho^{-4}$ distribution \citep{2011CQGra..28l5023S}, where a network SNR $\rho \geq 12$ is used as a threshold for detectability. The mass parameters are converted into a chirp mass $M_{c}$ given by
\begin{equation}
M_{c} = \frac{(m_{1}m_{2})^{3/5}}{(m_{1}+m_{2})^{1/5}},
\end{equation}
and a symmetric mass ratio $\eta$ given by
\begin{equation}
\eta = \frac{m_{1}m_{2}}{(m_{1}+m_{2})^2}.
\end{equation}
We then calculate the posteriors for the masses using the method in \citet{2017MNRAS.465.3254M}. The posteriors are generated in the chirp mass parameter space as
\begin{equation}
M_{c} = M_{c}^{T} \left[1 + \alpha \frac{12}{\rho} (r_{0} + r) \right]
\end{equation}
and the symmetric mass ratio as
\begin{equation}
\eta = \eta^{T} \left[1+ 0.03\frac{12}{\rho} (r_{0} + r) \right]
\end{equation}
Here $M_{c}^{T}$ and $\eta^{T}$ are the true values, $r_{0}$ is a random number drawn from a standard normal distribution that represents the shift of the mean of the posterior with respect to the true value, and $r$ is an array of random numbers from the standard normal distribution. As in \citet{2017MNRAS.465.3254M}, the parameter $\alpha$, which determines the width of the posterior distribution, is motivated by the previous studies of \citet{2015ApJ...807L..24L, 2015MNRAS.450L..85M} and has values of 0.01, 0.03, and 0.1 when $\eta^{T}>0.1$, $0.1>\eta^{T}>0.05$, and $0.05>\eta^{T}$, respectively. The posterior samples on $M_{c}$ and $\eta$ are then converted to samples back in the $m_1$ and $m_2$ parameter space. 

We then generate the spin posteriors using the method of \citet{2017MNRAS.471.2801S}. As for the mass posteriors, we sample values for $\chi_1$ and $\chi_2$ from the relevant distributions from each of the five sub-populations. Those values can then be used to determine the effective spin $\chi_{\mathrm{eff}}$ \citep{PhysRevD.64.124013, PhysRevD.78.044021, PhysRevLett.106.241101} defined as,
\begin{equation}
\chi_{\mathrm{eff}} = \frac{a_{1}\cos{\theta_{1}}+qa_{2}\cos{\theta_{2}}}{(1+q)}
\end{equation}
where the mass ratio $q=m_{2}/m_{1}$. The posteriors are generated in the $\chi_{\mathrm{eff}}$ and $\chi_{1}$ parameter space as,
\begin{equation}
\chi_{\mathrm{eff}} = \chi_{\mathrm{eff}}^{T} + \left[\beta \frac{12}{\rho} (r_{0} + r)  \right]
\end{equation}
\begin{equation}
\chi_{1} = \chi_{1}^{T} + \left[\gamma \frac{12}{\rho} (r_{0} + r)  \right]
\end{equation}
where $\beta=0.1$ and $\gamma=0.2$, as in \citet{2017MNRAS.471.2801S}, and limits of -1 and 1 are applied to both parameters. A posterior for the $\chi_{2}$ parameter is then determined from the posteriors for $\chi_{\mathrm{eff}}$, $\chi_{1}$ and q.  The same number of posterior samples, between 200 and 1000, are used for the spin parameters of a given event. 

The 90\% confidence interval contours for five representative posterior distributions from each of the three binary black hole sub-populations are shown in Figure \ref{fig:posts}. The mass posteriors show the typically expected banana shape. Their size is related to the SNR of the signal. The aligned spin posteriors are much wider than the mass posteriors, as spin is poorly constrained by gravitational-wave detections.

In a generic binary black hole, the black hole spins will not be perfectly aligned with the orbital angular momentum. In these systems, the black hole spins will precess \citep{1994PhRvD..49.6274A,2014PhRvD..89l4025G}. This causes the distribution of spin-orbit misalignment angles $\cos{\theta_{1,2}}$ to vary with orbital frequency (or equivalently, gravitational-wave frequency). The distribution of $\chi_\mathrm{eff}$ is approximately constant at the 2nd post-Newtonian order \citep{PhysRevD.78.044021}. An isotropic distribution of spins is expected to remain isotropic through post-Newtonian evolution \citep{2004PhRvD..70l4020S,2007ApJ...661L.147B}. These considerations will be important in choosing the parameterisation for clustering on real gravitational-wave observations.

\section{Clustering Method}
\label{sec:method}

In this study, we consider a Gaussian mixture model as implemented in Scikit learn \citep{scikit-learn}. A Gaussian mixture model fits a linear combination of multivariate Gaussian distributions to the input data. 
Our implementation of this method requires smooth inputs for clustering, so it is not possible to use all of the posterior samples as input.  We therefore represent each observation by a set of estimators: the median and 90\% confidence limit values of each marginalised one-dimensional posterior.  Thus, the input data $x$ for an observation with $n$ inferred parameters is a vector of length  $3\times n$.  An $n$-dimensional Gaussian probability density is given by
\begin{equation}
p(x|\mu,\Sigma) = \frac{1}{\sqrt{(2\pi)^n \mathrm{det} \Sigma }} \exp \left[ -\frac{1}{2} (x-\mu)^{T} \Sigma^{-1} (x-\mu)  \right] \, ,
\end{equation}
where $\mu$ is the mean, and $\Sigma$ is the covariance matrix. The likelihood function for a single observation under a $K$-component Gaussian mixture model is then given by
\begin{equation}
\mathrm{GMM}(x|w,\{\mu,\Sigma\}) = \sum_{k=1}^{K} w_{k} p(x|\mu_k,\Sigma_k) \, ,
\end{equation}
where $w_k$ are the mixture weights equal to marginal probabilities of mixture components. The free parameters for each Gaussian are its weight (up to an overall normalisation), mean and covariance matrix. Therefore, we need to find the optimum parameters that maximise the likelihood in order to fit the mixture of Gaussians to the observed compact binary population. This likelihood for the full set of $N$ observations  $X = \{X_1 ... X_N\}$ is given by
\begin{equation}
p(X|w,\mu,\Sigma) = \prod_{j=1}^{N} \mathrm{GMM}(X_{j}|w,\mu,\Sigma) \, .
\end{equation}
An expectation maximisation technique \citep{10.2307/2984875} is then used to find a maximum likelihood estimate that determines the correct values of the means, weights and covariance matrices for a given number of Gaussians. In this method, the weights, means and covariances are first randomly initialised. Then the expectation and maximisation steps are iterated repeatedly so that the likelihood of the data increases at the end of each step.

To determine the correct number of Gaussians we minimize the Bayesian Information Criterion (BIC) given by 
\begin{equation}
\mathrm{BIC} = -2 \ln{\mathcal{L}} + k \ln{N}\,
\end{equation}
where $k$ is the number of free parameters to be estimated, and $\mathcal{L}$ is the maximised value of the likelihood of the best fit model. BIC adds a penalty to models with larger numbers of parameters to avoid over fitting. The maximum number of Gaussians we consider for this method is 10. This method can produce results in a few seconds for very large numbers of detected compact binaries. Other mixture model methods that can be used for model selection include infinite mixtures and bottom-up growing and top-down pruning techniques \citep{NIPS1999_1745, Figueiredo:2002:ULF:507475.507482, 2018MNRAS.479..601D}.

\section{Results}
\label{sec:results}

\begin{table*}
\begin{center}
\begin{tabular}{|c|c|c|c|c|c|c|}
\hline
 Parameters & No. predicted & BNS & NSBH & BBH1 & BBH2 & BBH3 \\   
 & sub-populations & & & & & \\ \hline
 $m_1$, $m_2$      & 4 & 98\% in the & 99\% in the & mixed with BBH2 & mixed with BBH 1 & 79\% in the \\ 
 & & same cluster & same cluster & & & same cluster \\ \hline
 $\chi_1$, $\chi_2$  & 4 & mixed with BBH1 & 95\% in the &  mixed with BNS & 78\% in the & 57\% in the \\ 
 & & & same cluster & & same cluster & same cluster \\ \hline
 $m_1$, $m_2$, $\chi_1$, $\chi_2$ & 5 & 99\% in the & 100\% in the & 80\% in the & 68\% in the & 81\% in the \\ 
 & & same cluster & same cluster & same cluster & same cluster & same cluster \\ \hline
 $\log M_{c}$, $\chi_{\mathrm{eff}}$ & 5 & 92\% in the & 92\% in the & mixed with BBH2 & mixed with BBH1 & 91\% in the \\ 
 & & same cluster & same cluster & & & same cluster \\ \hline
 BBH only, $m_1$, $m_2$, $\chi_1$, $\chi_2$ & 3 & & & 74\% in the & 47\% in the & 68\% in the \\ 
 & & & & same cluster & same cluster & same cluster \\ \hline
 BBH only, $\log M_{c}$, $\chi_{\mathrm{eff}}$ & 3 & & & mixed with BBH2 & mixed with BBH1 & 92\% in the \\  
 & & & & & & same cluster \\ \hline
\end{tabular}
\caption{Gaussian mixture model results for different combinations of parameters. 
The best results are obtained when parametrising the full population with $m_1$, $m_2$, $\chi_1$, and $\chi_2$.}
\label{tab:gmm_results}
\end{center}
\end{table*} 

Here we report on the results of applying a Gaussian mixture model to our mock population. We are particularly interested in the number of clusters re-constructed (does it equal the five modelled sub-populations) and the weights assigned to each cluster (does it equal 20\% as in the model).  
We also track the fraction of events from the same sub-population that are clustered together.  These results are given in Table \ref{tab:gmm_results}.

We vary the choices of parameters used for clustering.  Clustering requires a choice of a metric on the parameter space in order to define a distance between clusters and a choice for the shape of each cluster.  Neither can be determined from first principles.  While the assumed Gaussian probability density with covariance matrices define the cluster shape and a (Euclidean) distance metric, changing the parametrisation effectively changes the cluster shape and distance; for example, transforming to logarithmic coordinates is equivalent to assuming a log-normal cluster shape rather than a normal one.
We thus consider using log chirp mass and the effective spin as clustering parameters, as these are better determined from observations than individual masses and spins.  We also include results using only mass, only spin, and only the binary black hole sub-populations. The results are given in Table \ref{tab:gmm_results}. 

Although optimality is difficult to define, we find that for our particular population of events, using individual masses and spins as clustering parameters yields robust results.  The number of sub-populations is generally correctly determined for all parameter options.  As expected, only 4 sub-populations are found when using the mass parameters alone, since BBH1 and BBH2 have identical mass distributions.  When using spins alone, similar results to the combined spin and mass case are found for sub-population BBH2, however only around half of the BBH3 binaries are in the same sub-population, with the other half mixed in with the other sub-populations.   Two of the sub-populations with zero spin, BNS and BBH1, were mixed as expected. However, the NSBH sub-population that also has zero spin was still distinguishable from the other zero spin sub-populations. This is due to the mass ratio value of the NSBH sub-population changing the shape of the posteriors in the individual spin parameter space.  Log chirp mass and effective spin also lead to BBH1 and BBH2 being mixed into two classes that are split into lower and higher masses, rather than by spins, despite their different spin distributions.  This may be partly related to a partial artificial breaking of the spin-degeneracy in our spin measurement uncertainty model.   

When using only the binaries in the three BBH sub-populations, the results show that there is a larger mixing between sub-populations than when the BNS and NSBH sub-populations are included.  Using all possible compact binaries and the full set of parameters $m_1$, $m_2$, $\chi_1$, and $\chi_2$ yields the most information, and the Gaussian mixture model performs better with a larger set of input data. Now that the best parameterisation has been identified, the rest of the results in this section use only $\chi_{1}$, $\chi_{2}$, $m_{1}$ and $m_{2}$ as input parameters. 

\begin{figure*}
\centering
\includegraphics[width=0.45\textwidth]{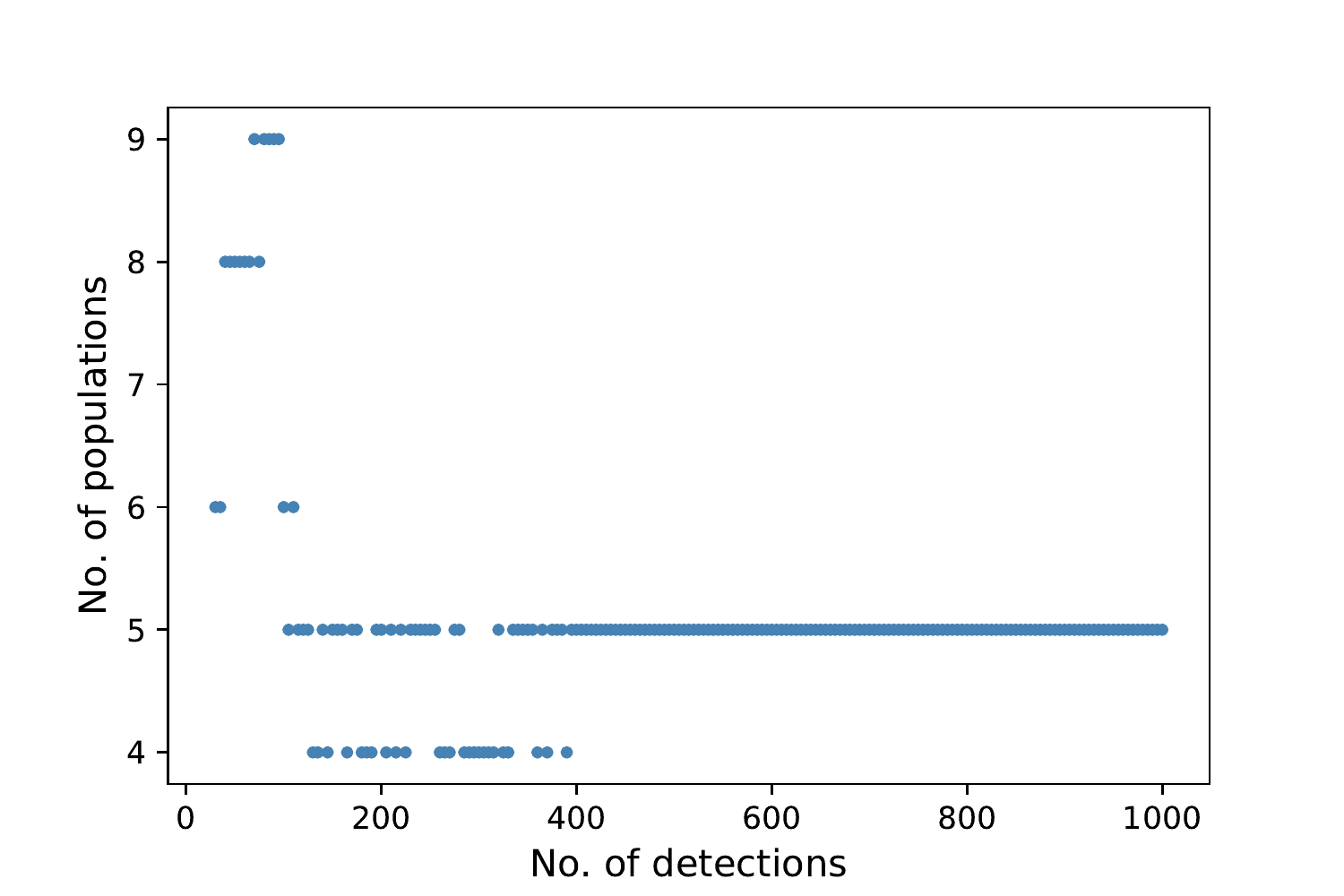}
\includegraphics[width=0.45\textwidth]{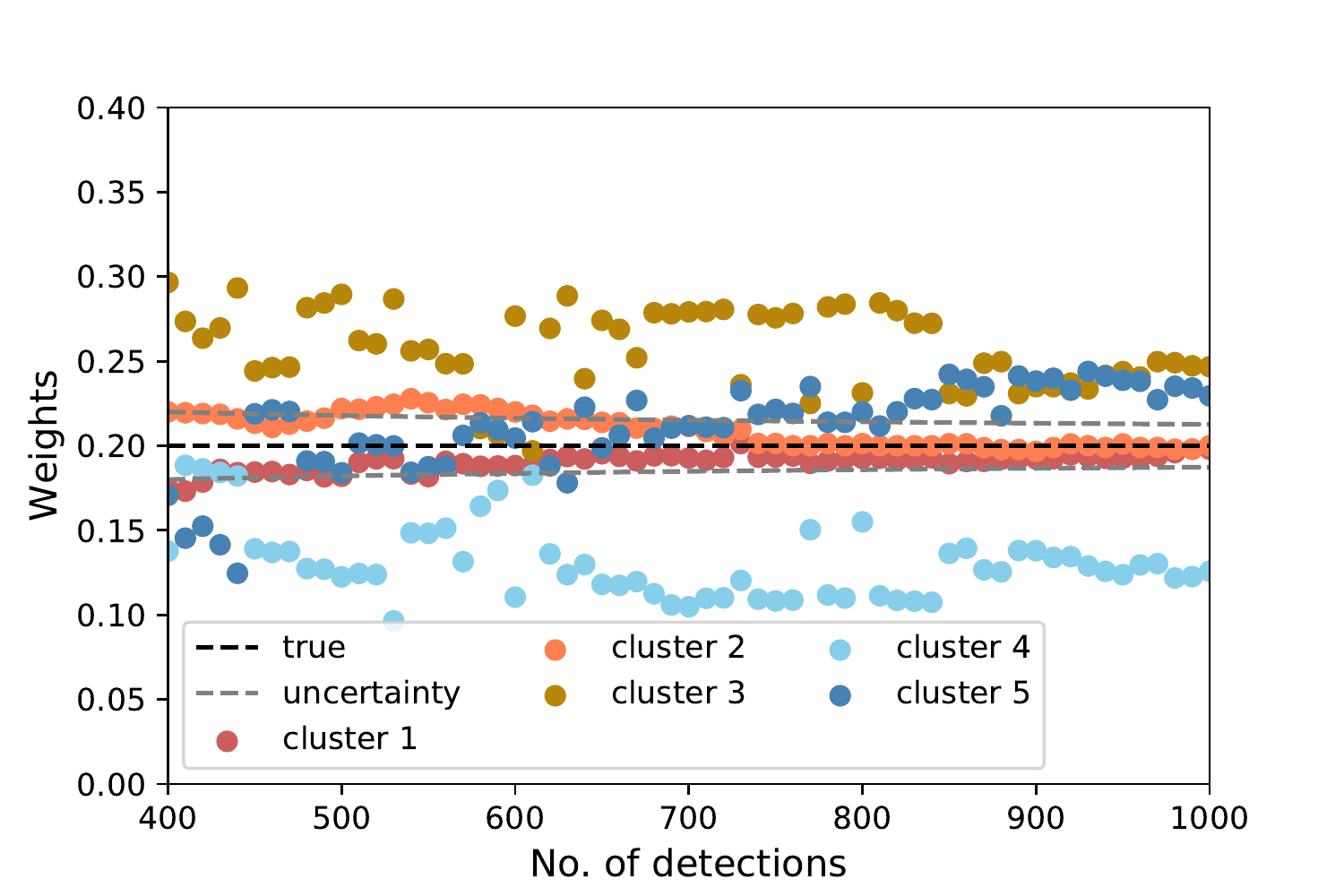}
\caption{(left) The number of sub-populations estimated by the Gaussian mixture model parametrised with $\chi_{1}$, $\chi_{2}$, $m_{1}$ and $m_{2}$ as the number of detections is increased. Approximately 400 detections are needed to estimate the correct number of sub-populations. (right) The estimated weights of the five clusters after the correct number of clusters is determined.}
\label{fig:gmm_nodet}
\end{figure*}

We investigate how many detections are needed before the correct number of sub-populations can be identified. We detect the binary signals in a random order starting with 30 detections, increased in steps of five detections, and apply the Gaussian mixture model after each step. The results are shown in Figure \ref{fig:gmm_nodet}. With only 30 detections, the smallest number we consider, we can distinguish between the BNS, NSBH, and BBH sub-populations. The three BBH sub-populations are then split into a large number of Gaussians by the mass of the binaries. This is because the masses are  better constrained by gravitational-wave signatures than the spin, therefore a larger number of spin posteriors are required before the spin has a significant influence on the results. 
Accidental apparent clustering of a small number of observations in a multi-dimensional space leads to an over-estimate of the number of clusters for $\lesssim 100$ detections. 
After 105 detections, the binaries in the BBH3 sub-population are correctly grouped together by the Gaussian mixture model, but the BBH1 and BBH2 sub-populations are still mixed either into the same sub-population or into two sub-populations that are split by their mass. After 400 detections we can start to distinguish between the more difficult case of the BBH1 and BBH2 sub-populations. 

In Figure \ref{fig:gmm_nodet} we also show the weight values after the correct number of sub-populations has been determined. The correct number for all sub-populations should be 0.2 with a multinomial counting uncertainty that is given by 
\begin{equation}
\mathrm{uncertainty} = \sqrt{ \frac{N_{\mathrm{pop}}-1}{N_{\mathrm{pop}}^2 N_{\mathrm{det}}} } = 0.4 N_{\mathrm{det}}^{-0.5},
\end{equation}
where $N_{\mathrm{pop}}=5$ is the number of sub-populations, and $N_{\mathrm{det}}$ is the number of detections.
The BNS and NSBH sub-populations have the correct weight values. There is a larger error in the weights of the three BBH sub-populations due to some mixing between the different sub-populations. This occurs due to the poor gravitational-wave spin measurements and two of the sub-populations only differing in their spin values. 

\begin{figure*}
\centering
\includegraphics[width=0.48\textwidth]{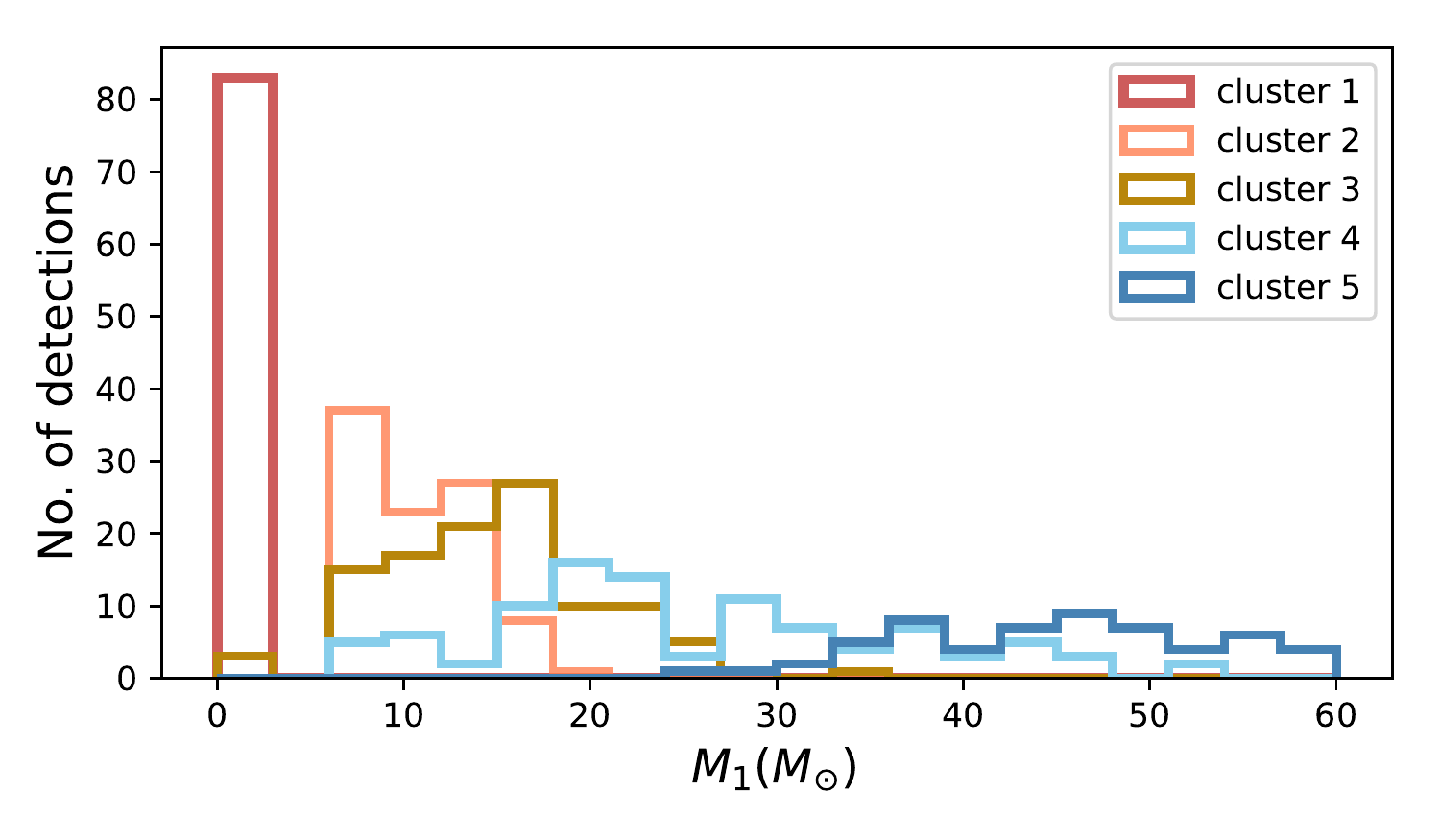}
\includegraphics[width=0.48\textwidth]{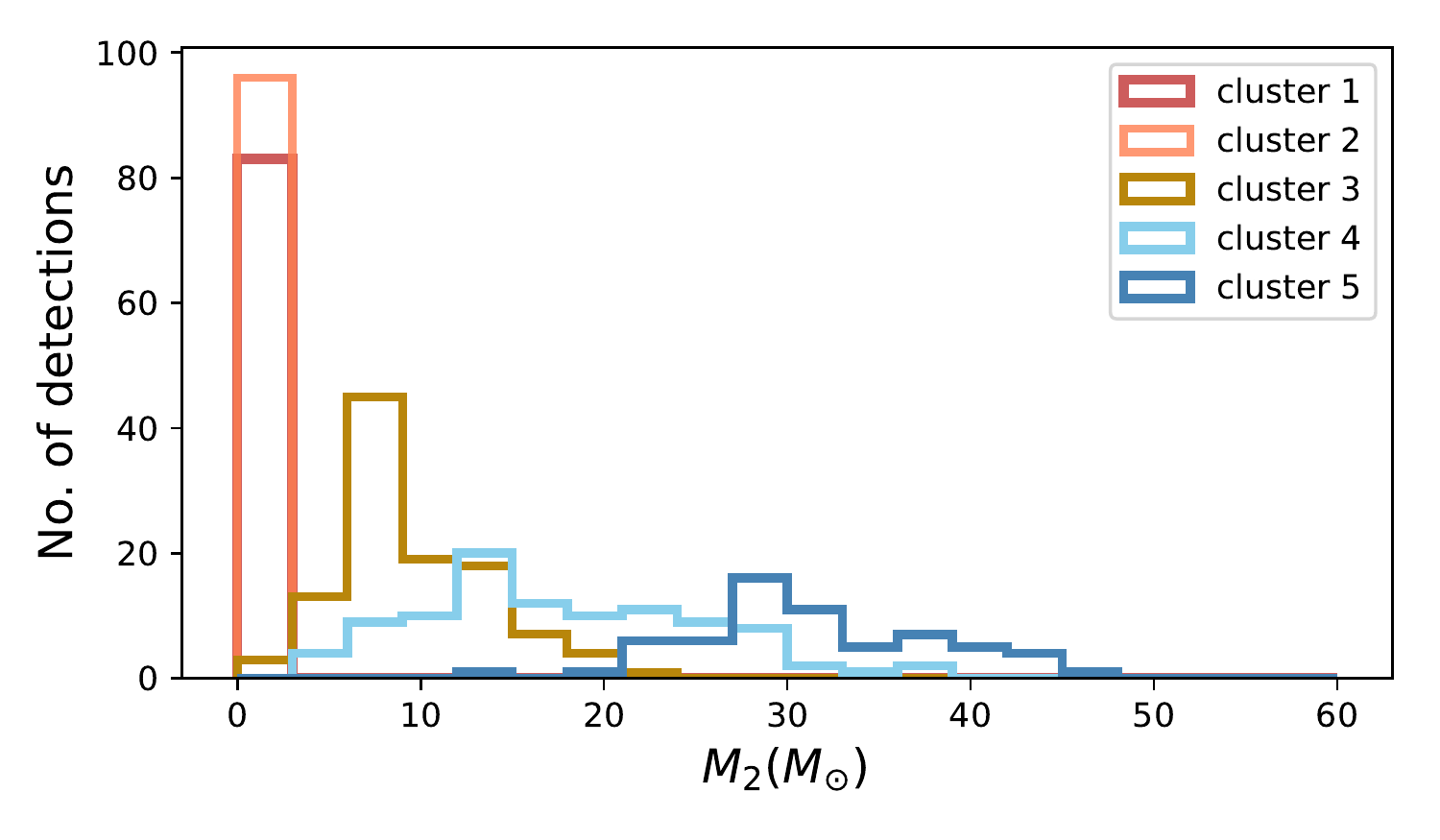}
\includegraphics[width=0.48\textwidth]{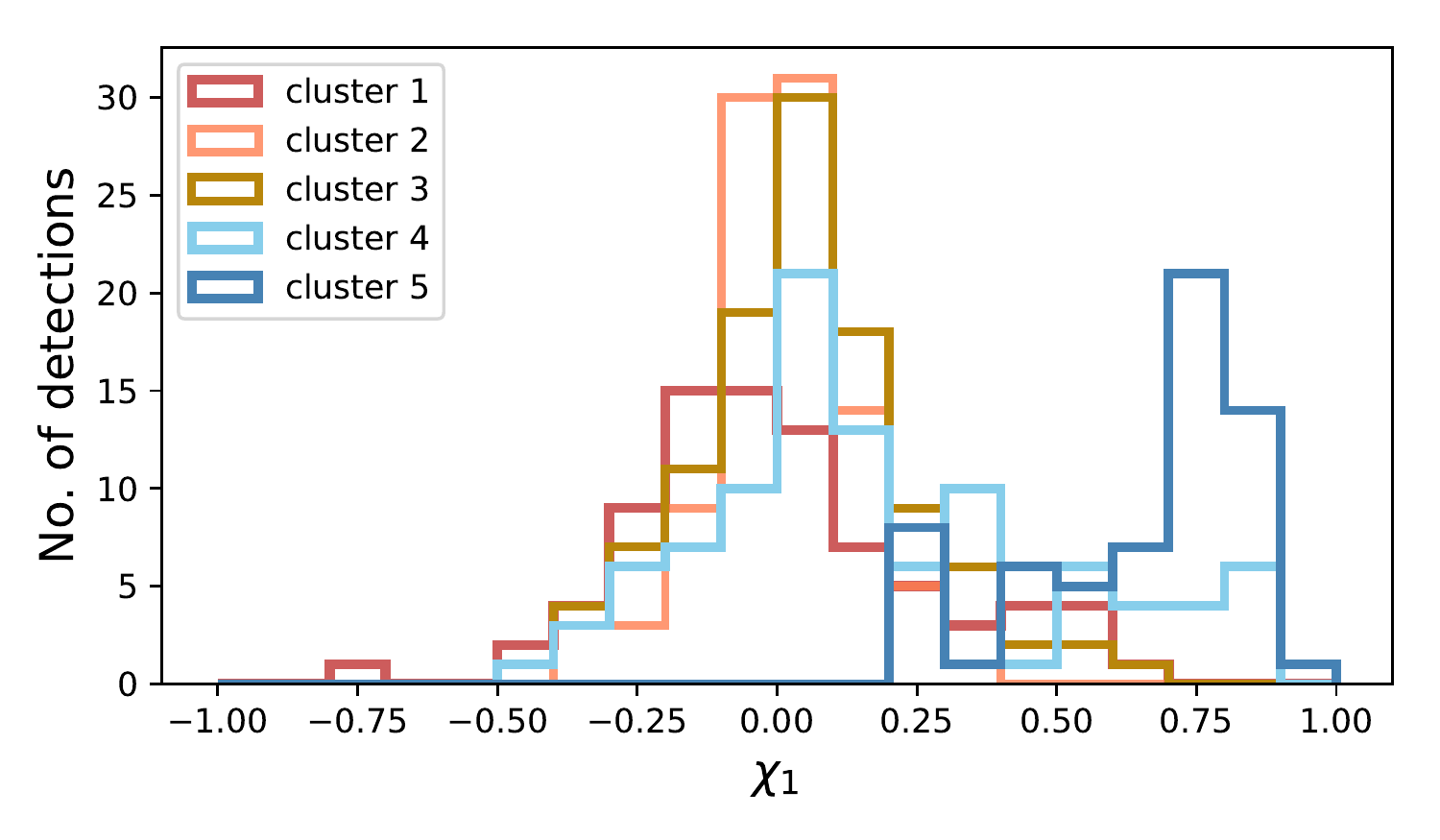}
\includegraphics[width=0.48\textwidth]{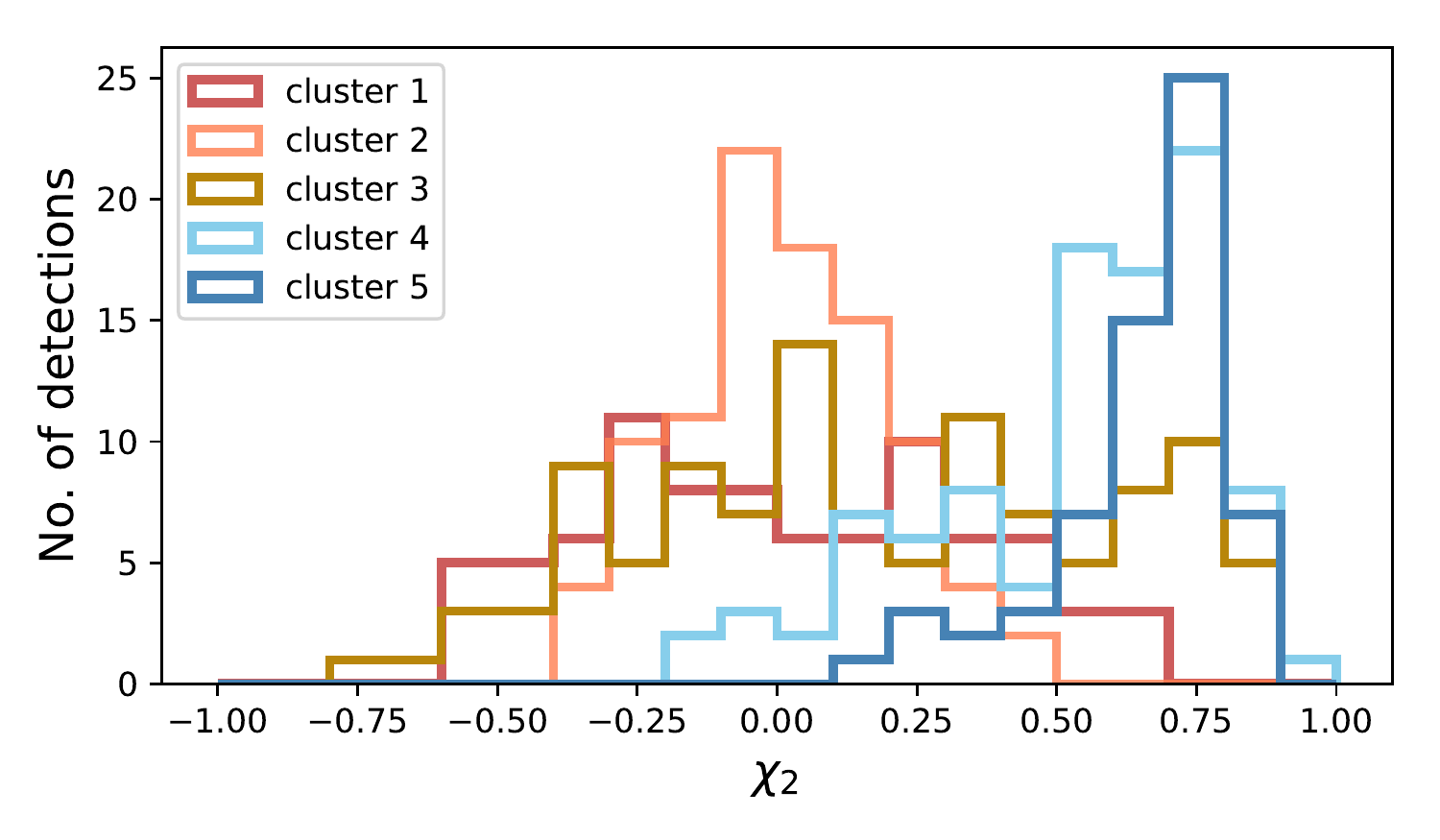}
\caption{The individual mass and spin distributions (binned posterior medians) of the sub-populations identified by the Gaussian mixture model after 400 detections. (top left) The five $m_{1}$ distributions. (top right) The $m_{2}$ distributions. (bottom left) The $\chi_{1}$ distributions. Four of the distributions are clearly centred on zero, whilst one distribution has higher values. (bottom right) The $\chi_{2}$ distributions. Two of the distributions have higher $\chi_{2}$ values than the other three, which are centred on zero. }
\label{fig:gmm_dists}
\end{figure*}

After finding the correct number of sub-populations, we want to know the mass and spin distributions of each individual sub-population, as this will aid investigations into differences in their formation mechanisms. The individual mass and spin distributions determined by this method after 400 detections are shown in Figure \ref{fig:gmm_dists}. 

In cluster 1, associated with the BNS sub-population, both masses have distributions expected for neutron stars and both of the aligned spin distributions are centred on zero. In cluster 2, associated with the NSBH sub-population, mass distributions show that $m_{1}$ is typically a low mass black hole and that $m_{2}$ is a neutron star. The spin distributions of cluster 2 (NSBH) and cluster 3 (BBH1) are centred on zero, as expected. Cluster 5, associated with the BBH3 sub-population, contains systems with high masses and high spins, while cluster 4 (BBH2)  exhibits a $\chi_{2}$ distribution favouring high spin while the $\chi_{1}$ distribution is centred on zero. 

We investigate the impact of different mixing ratios in the mock population by considering two other population models consisting of the same five sub-populations present with different relative mixing ratios. In all cases, we evaluate the clustering results after 620 detections. The first new mixture, which we refer to as mixture 2, consists of a much smaller number of BNS and NSBH signals as expected from current gravitational-wave detections. The total population in this mixture consists of 6.5\% BNS, 6.5\% NSBH, 29\% BBH1, 29\% BBH2, and 29\% BBH3 signals. The second new mixture, which we refer to as mixture 3, has a smaller number of BNS, NSBH, and BBH in one of the BBH sub-populations. This mixture consists of 16\% BNS, 16\% NSBH, 16\% BBH3, 26\% BBH1, and 26\% BBH2 signals. 

The results for the different mixtures are shown in Table \ref{tab:gmm_ratios}. Our clustering algorithm determines the correct number of sub-populations for all of the mixing ratios considered. For mixture 2, where there is now a much smaller number of BNS and NSBH detections, the increase in the ratio of detections from all three BBH sub-populations leads to an improved clustering result for those three sub-populations. For mixture 3, where we also decrease the BBH3 population, the accuracy of the results is similar to the equal mixture case.

\begin{table*}
\begin{center}
\begin{tabular}{|c|c|c|c|c|c|c|}
\hline
 Mixtures & No. predicted & BNS & NSBH & BBH1 & BBH2 & BBH3 \\   \hline
 equal ratio & 5 & 99\% in the & 100\% in the & 86\% in the & 69\% in the & 64\% in the \\ 
 & & same cluster & same cluster & same cluster & same cluster & same cluster \\ \hline
 mixture 2 & 5 & 100\% in the & 100\% in the & 81\% in the & 79\% in the & 91\% in the \\ 
 & & same cluster & same cluster & same cluster & same cluster & same cluster \\ \hline
 mixture 3 & 5 & 99\% in the & 100\% in the & 79\% in the & 79\% in the & 65\% in the \\ 
 & & same cluster & same cluster & same cluster & same cluster & same cluster \\ \hline
\end{tabular}
\caption{The results after 620 detections when analysing populations consisting of a mixture of the same five sub-populations with different mixing ratios. Mixture 2 contains a much smaller number of BNS and NSBH. Mixture 3 has fewer detections in the BBH3 sub-population as well BNS and NSBH. The Gaussian mixture model isolates 5 sub-populations for all mixtures considered.}
\label{tab:gmm_ratios}
\end{center}
\end{table*}

\section{Conclusions}
\label{sec:conclusion}

Measuring the number, properties, and shapes of the sub-populations associated with different source types and formation channels of merging compact binaries detected through their gravitational-wave signatures will assist in the astrophysical interpretation of these observations.  We explore clustering on simulated distributions for the masses and spins of binary systems from a mixture of five sub-populations, using mock inference on their individual parameters from gravitational-wave observations.  
We demonstrate that Gaussian mixture model clustering on the full set of observations in the four-dimensional space of source parameters makes it possible to distinguish the multiple sub-populations present after 400 detections and determine the mass and spin distributions of each sub-population. 

Gaussian mixture model clustering is robust and computationally efficient.  It scales well to higher-dimensional parameter spaces, but it requires a partial loss of the posterior information, as only a few estimators rather than the full event posteriors are used for clustering. On the other hand, as with any clustering technique, it relies on the choice of a distance metric in the parameter space. It also assumes a specific parametrised shape for the clusters in the mixture model.  Although the method performs robustly on the mock populations we simulate, which are not multi-variate Gaussians, it may be suboptimal for some non-Gaussian distribution shapes. This method requires an upper limit on the number of sub-populations; we use the BIC to determine the optimal number of sub-populations. Finally, this method does not allow us to produce uncertainty estimates on the number of sub-populations and their weights; however, this is a general challenge for any clustering method that relies on an arbitrary distance metric, and is not a specific flaw of the current technique.

The difficulty in distinguishing sub-populations depends on the similarity of the parameters of the binaries in the different sub-populations, and the exact number of observations necessary for this therefore depends on the population properties. The most difficult case we consider here is two sub-populations of binary black holes which have identical mass distributions and differ only in one of their spin components.  We do not incorporate selection effects in our analysis \citep{2015PhRvD..91b3005F, 2018arXiv180902063M}.   We also do not consider the effect of lower significance gravitational-wave detections, which may be produced by transient noise that could not be distinguished from gravitational-wave sources \citep{2018arXiv180903815G}.

\section*{Acknowledgments}

JP, SS, and IM are supported by the Australian Research Council Centre of Excellence for Gravitational Wave Discovery (OzGrav), through project number CE170100004. PT was supported by the European Commission Horizon 2020 Innovative Training Network SUNDIAL (SUrvey Network for Deep Imaging Analysis and Learning), Project ID: 721463.


\bibliographystyle{mnras}
\interlinepenalty=20000
\bibliography{bibfile} 





\bsp	
\label{lastpage}
\end{document}